\documentclass{ws-procs975x65}            % default, citations in superscript
\usepackage{latexsym}
\usepackage{revsymb}
\usepackage{subfigure}
\begin{document}
%\markboth{Chang, Minic, Roman, Sun, Takeuchi}
%{On the physics of the minimal length : the question of gauge invariance}

\newcommand{\bra}[1]{\langle #1 |}
\newcommand{\ket}[1]{| #1 \rangle}
\newcommand{\vev}[1]{\left\langle #1 \right\rangle}
\newcommand{\bvev}[1]{\bigl\langle #1 \bigr\rangle}
%\newcommand{\Bvev}[1]{\Bigl\langle #1 \Bigr\rangle}

%%%%%%%%%%%%%%%%%%%%%%%%%%%%%%%%%%%%%%%%%%%%%%%%%%%%%%%%%%%%%%%%%%%%%%%%%%%%%%%%%%%%%%%%%%%%%%%%%
%%%%%%%%%%%%%%%%%%%%%%%%%%%%%%%%%%%%%%%%%%%%%%%%%%%%%%%%%%%%%%%%%%%%%%%%%%%%%%%%%%%%%%%%%%%%%%%%%

\title{On the Physics of the Minimal Length:\\The Question of Gauge Invariance}

\author{Lay Nam Chang${}^{1}$, Djordje Minic${}^{1}$, Ahmed Roman${}^{1,2}$, Chen Sun${}^{1}$, and Tatsu Takeuchi${}^{1}$}
\address{${}^{1}$Center for Neutrino Physics, Department of Physics, Virginia Tech, Blacksburg VA 24061 USA\\
${}^{2}$Department of Organismic and Evolutionary Biology, Harvard University, Cambridge MA 02138, USA\\
laynam@vt.edu, dminic@vt.edu, mido@vt.edu, sunchen@vt.edu, takeuchi@vt.edu
}

\begin{abstract}
In this note we discuss the question of gauge invariance in the
presence of a minimal length. This contribution is prepared for the celebration of 
the 60th anniversary of the Yang-Mills theory.
\end{abstract}


\bodymatter
%%%%%%%%%%%%%%%%%%%%%%%%%%%%%%%%%%%%%%%%%%%%%%%%%%%%%%%%%%%%%%%%%%%%%%%%%%%%%%%%%%%%%%%%%%%%%%%%%
%%%%%%%%%%%%%%%%%%%%%%%%%%%%%%%%%%%%%%%%%%%%%%%%%%%%%%%%%%%%%%%%%%%%%%%%%%%%%%%%%%%%%%%%%%%%%%%%%
\section{Introduction}

Yang-Mills theory \cite{Yang:1954ek} 
represents one of the most remarkable achievements of theoretical physics
in the second half of the 20th century.\footnote{For a historical account of the concept of gauge invariance that lead to Yang-Mills theory, see Ref.~\citenum{O'Raifeartaigh:1997ia}.
See also the recent account by Yang in Ref.~\citenum{Yang:2014}.} 
Its fundamental importance is found in high energy physics as well as condensed matter physics and mathematics.\footnote{For a recent discussion of the mass gap problem in the pure Yang-Mills theory\cite{Jaffe:2000ne} see, for example,
Refs.~\citenum{Karabali:1995ps,Karabali:1998yq,Leigh:2005dg,Leigh:2006vg,Freidel:2006qy,Freidel:2008pw}. 
}
In this note we address the issue of gauge invariance which underlies the structure of Yang-Mills theory in the context of theories endowed with a minimal length.

One of the ubiquitous features of quantum gravity\cite{Kiefer:2012boa} is its
possession of a fundamental length scale.  In a theory of quantum gravity such as string theory \cite{Green:1987sp,Green:1987mn,Polchinski:1998rq,Polchinski:1998rr,Becker:2007zj},
the fundamental scale determines the typical spacetime extension of a fundamental string \cite{Chang:2011jj}.
In canonical string theory this is $\ell_S = \sqrt{\alpha'}$, where $\hbar c/\alpha'$ is the string tension.
Such a feature is to be expected of any candidate theory of quantum gravity,
since gravity itself is characterized by the Planck length $\ell_P = \sqrt{\hbar G_N/c^3}$.
Moreover, $\ell_P\sim\ell_S$ is understood to be the 
\textit{minimal length} below which spacetime distances cannot be resolved \cite{Wheeler:1957mu,Mead:1964zz,Maggiore:1993rv,Garay:1994en,Hossenfelder:2012jw}:
\begin{equation}
\delta s \;\agt\; \ell_P \sim \ell_S\;.
\label{MinimalLength}
\end{equation}
Local quantum field theory, on the other hand, 
is completely oblivious to the presence of such a scale,
despite being the putative infrared limit of some more fundamental theory, such as string theory.
A natural question to ask is, therefore, whether the formalism of quantum mechanics (and thus of an effective local quantum field theory) can be deformed or
extended in such a way as to consistently incorporate the minimal length.  
If it is at all possible, the precise manner in which quantum mechanics must be modified
may point to solutions of yet unresolved mysteries such as the cosmological constant problem\cite{Weinberg:1988cp,Carroll:2000fy,Witten:2000zk,Straumann:2002tv,Nobbenhuis:2004wn},
which is essentially quantum gravitational in its origin. 
Such a structure should also illuminate the nature of quantum gravity, including string theory\cite{Polchinski:1994mb,Freidel:2013zga,Freidel:2014qna,Freidel:2015pka,Freidel:2015uug}.

The very idea of the minimal length in the context of quantum theory has a fascinating and long history.
It was used by Heisenberg in 1930\cite{Kragh:1995,Carazza:1995} to address
the infinities of the then newly formulated theory of quantum electrodynamics\cite{Heisenberg:1929xj}.
Over the years, the idea has been picked up by many authors in a plethora of contexts, e.g. 
Refs.~\citenum{born:1933,Snyder:1946qz,Snyder:1947nq,Yang:1947ud,Mead:1966zz,Padmanabhan:1986ny,Padmanabhan:1987au,Padmanabhan:1996ap,Kato:1990bd} to list just a few.
(For a comprehensive bibliography consult Ref.~\citenum{Hossenfelder:2012jw}.)
Various ways to deform or extend quantum mechanics have also been suggested\cite{Weinberg:1989cm,Weinberg:1989us,Bender:2002vv}.
In this note, we focus our attention on how a minimal length can be introduced into quantum mechanics
by modifying its algebraic structure \cite{Maggiore:1993zu,Maggiore:1993kv,Kempf:1994su}. In particular, we follow our previous work in Refs.~\citenum{Chang:2001kn,Chang:2001bm,Benczik:2002tt,Benczik:2002px,Benczik:2005bh,Chang:2010ir,Lewis:2011fg,Lewis:2014wfa}.

%%%%%%%%%%%%%%%%%%%%%%%%%%%%%%%%%%%%%%%%%%%%%%%%%%%%%%%%%%%%%%%%%%%%%%%%%%%%%%%%%%%%%%%%%%%%%%%%%

The starting point of our discussion is the minimal length
uncertainty relation (MLUR) \cite{Amati:1988tn,Witten:2001ib}, also known as the
generalized uncertainty principle (GUP)
\begin{equation}
\delta x \;\sim\;
\left(\dfrac{\hbar}{\delta p} + \alpha'\,\dfrac{\delta p}{\hbar}\right)\;,
\label{MLUR}
\end{equation}
which is suggested by a re-summed
perturbation expansion of the string-string scattering amplitude in a flat spacetime background \cite{Gross:1987kza,Gross:1987ar,Amati:1987wq,Amati:1987uf}.
This is essentially a Heisenberg microscope argument \cite{Heisenberg:1930}
in the S-matrix language\cite{Wheeler:1937zz,Heisenberg:1943a,Heisenberg:1943b,Heisenberg:1944} with fundamental strings used to probe fundamental strings.
The first term inside the parentheses on the right-hand side is the usual
Heisenberg term coming from the shortening of the probe-wavelength as momentum is increased,
while the second-term can be understood as due to the lengthening of the probe string as more energy is pumped into it:
\begin{equation}
\delta p 
\;=\; \dfrac{\delta E}{c} 
\;\sim\; \dfrac{\hbar}{\alpha'}\,\delta x 
%\;=\; \dfrac{\hbar}{\ell_S^2}\,\delta x 
%\;=\; \dfrac{1}{\hbar\beta}\,\delta x
\;.
\label{dPproptodX}
\end{equation}
Eq.~(\ref{MLUR}) implies that the uncertainty in position, $\delta x$, is bounded from 
below by the string length scale,
\begin{equation}
\delta x \;\agt\; \sqrt{\alpha'} \;=\; \ell_S \;, 
\label{dXmin}
\end{equation}
where the minimum occurs at 
\begin{equation}
\delta p \;\sim\; \dfrac{\hbar}{\sqrt{\alpha'}} \;=\; \dfrac{\hbar}{\ell_S} \;\equiv\; \mu_S\;.
\end{equation}
Thus, $\ell_S$ is the minimal length below which spatial distances cannot be resolved,
consistent with Eq.~(\ref{MinimalLength}).
In fact, the MLUR can be motivated by fairly elementary general relativistic considerations 
independent of string theory, which suggests that it is a universal feature of quantum gravity \cite{Wheeler:1957mu,Mead:1964zz,Maggiore:1993rv,Garay:1994en}.

Note that in the trans-Planckian momentum region $\delta p \gg \mu_S$, the MLUR is
dominated by the behavior of Eq.~(\ref{dPproptodX}), which implies that large $\delta p$ (UV)
corresponds to large $\delta x$ (IR), and that there exists a correspondence
between UV and IR physics.
Such UV/IR relations have been observed in various string dualities\cite{Green:1987sp,Green:1987mn,Polchinski:1998rq,Polchinski:1998rr,Becker:2007zj},
and in the context of AdS/CFT correspondence\cite{Susskind:1998dq,Peet:1998wn}
(albeit between the bulk and boundary theories).
Thus, the MLUR captures another distinguishing feature of string theory.\footnote{In addition to the MLUR, 
another uncertainty relation has been pointed out by Yoneya as characteristic of string theory.
This is the so-called spacetime uncertainty relation (STUR)
$
\delta x\,\delta t \;\sim\; \ell_S^2/c\;,
\label{STUR}
$
which can be motivated in a somewhat hand-waving manner by combining the usual energy-time uncertainty relation $\delta E\,\delta t\sim \hbar$ with Eq.~(\ref{dPproptodX}).\cite{Mandelshtam:1945,Wigner:1972,Bauer:1978wd} 
However, it can also be supported via an analysis of D0-brane scattering in certain backgrounds 
in which $\delta x$ can be made arbitrary small 
at the expense of making the duration of the interaction $\delta t$ arbitrary large\cite{Yoneya:1989ai,Yoneya:1997gs,Yoneya:1997dv,Yoneya:2000bt,Yoneya:2000sf,Li:1996rp,Li:1998fc,Jevicki:1998yr,Awata:1999dz,Minic:1998nu}.
While the MLUR pertains to dynamics of a particle in a non-dynamic spacetime, the STUR
can be interpreted to pertain to the dynamics of spacetime itself in which 
the size of a quantized spacetime cell is preserved.}

Note that the recent reformulation of quantum gravity in the guise of metastring theory\cite{Freidel:2013zga,Freidel:2014qna,Freidel:2015pka,Freidel:2015uug} based on such novel concepts as
relative locality\cite{AmelinoCamelia:2011bm} and dynamical energy-momentum space,\footnote{There are many examples of fundamental physics structures that stay unchanged under the interchange of conjugate variables, such as spatial and momentum coordinates: $x \to p$ and $p \to - x$. 
See Refs.~\citenum{Born:1935,Born:1949}.
By combining the uncertainty relation between $x$ and $p$ and the fact that $x$ is endowed with a metric structure in general relativity, Born reciprocity would suggest a dynamical momentum and thus a dynamical phase space. See Ref.~\citenum{Born:1938}.} 
strongly suggests that the fundamental length scale always goes hand-in-hand with the
fundamental energy scale, and that the two appear in a symmetric manner, 
suggested by Born reciprocity.\footnote{%
Some of the historically interesting follow-up work on Born reciprocity and dynamical energy-momentum space can be found in Refs.~\citenum{Yukawa:1950eq,Golfand:1960,Golfand:1963a,Golfand:1963b,Veneziano:1986zf,Batalin:1989dm,Freidel:2005me,Bars:2010xi}.
}
In this note, we concentrate only on the implications of the fundamental length scale.
In particular, we make a few observations on the question of gauge invariance in the presence of a minimal length. This seems to be an appropriate topic for the
celebration of the 60th anniversary of Yang-Mills theory\cite{Yang:1954ek}.

%%%%%%%%%%%%%%%%%%%%%%%%%%%%%%%%%%%%%%%%%%%%%%%%%%%%%%%%%%%%%%%%%%%%%%%%%%%%%%%%%%%%%%%%%%%%%%%%%
\section{Local Gauge Invariance in the Presence of a Minimal Length}

The question of gauge invariance in the presence of a minimal length has been addressed in the context of non-commutative field theories (NCFT) \cite{Douglas:2001ba,Szabo:2001kg,Polychronakos:2007df,Szabo:2009tn,Blaschke:2010kw,D'Ascanio:2016asa,Borowiec:2016zrc}. In that case the minimal length can be related, in a particular realization, to an effective
magnetic length. 
Here we want to investigate this question from a different point of view, following our previous work on the minimal
length as motivated by the foundations of quantum gravity and string theory\cite{Chang:2011jj}.
As we will note in what follows, the algebra of NCFT, in which the question of gauge invariance is fully understood, is a limiting case of a more general discussion involving a minimal length motivated by quantum gravity and string theory.

%%%%%%%%%%%%%%%%%%%%%%%%%%%%%%%%%%%%%%%%%%%%%%%%%%%%%%%%%%%%%%%%%%%%%%%%%%%%%%%%%%%%%%%%%%%%%%%%%
\subsection{Commutation Relations}

To place the MLUR, Eq.~(\ref{MLUR}), onto a firmer footing, we begin by rewriting it as
\begin{equation}
\delta x\,\delta p \;\ge\; \dfrac{\hbar}{2}\left(1 + \beta\, {\delta p}^2\right)\;,
\label{MLUR2}
\end{equation}
where we have introduced the parameter $\beta = \alpha'/\hbar^2$.
%The minimum value of $\delta x$ as a function of $\delta p$ is plotted in Fig.~\ref{MLURfig}.
This uncertainty relation can be reproduced by deforming the canonical commutation 
relation between $\hat{x}$ and $\hat{p}$ to:
\begin{equation}
\dfrac{1}{i\hbar}\left[\,\hat{x},\,\hat{p}\,\right]\;=\; 1
\qquad\longrightarrow\qquad
\dfrac{1}{i\hbar}\left[\,\hat{x},\,\hat{p}\,\right]\;=\; A(\hat{p}^2)\;,
\label{MLUR3}
\end{equation}
with $A(p^2)=1+\beta p^2$. Indeed, we find
\begin{equation}
\delta x\,\delta p\;\ge\;
\dfrac{1}{2}\Bigl|\bvev{\left[\,\hat{x},\,\hat{p}\,\right]}\Bigr|
\;=\; \dfrac{\hbar}{2}
\left(1+\beta\vev{\hat{p}^2}\right)
\;\ge\; \dfrac{\hbar}{2}
\left(1+\beta\,\delta p^2\right)\;,
\end{equation}
since $\delta p^2 = \bvev{\hat{p}^2}-\bvev{\hat{p}}^2$.
The function $A(p^2)$ can actually be more generic, with $\beta p^2$ being the
linear term in its expansion in $p^2$.

When we have more than one spatial dimension, the above commutation relation can
be generalized to\cite{Chang:2011jj}
\begin{equation}
\dfrac{1}{i\hbar}[\,\hat{x}_i,\,\hat{p}_j\,]\;=\; A(\hat{\mathbf{p}}^2)\,\delta_{ij}+B(\hat{\mathbf{p}}^2)\,\hat{p}_i\hat{p}_j\;,
\label{deformedXP}
\end{equation}
where $\hat{\mathbf{p}}^2 = \sum_i \hat{p}_i^2$.
The right-hand side is the most general form that depends only on the momentum and 
respects rotational symmetry.  
Assuming that the components of the momentum commute among themselves,
\begin{equation}
[\,\hat{p}_i,\,\hat{p}_j\,] \;=\; 0\;,
\label{PP}
\end{equation}
the Jacobi identity demands that
\begin{equation}
\dfrac{1}{i\hbar}[\,\hat{x}_i,\,\hat{x}_j\,]
\;=\; -\left\{\,2(\hat{A}+\hat{B}\hat{\mathbf{p}}^2)\hat{A}' - \hat{A}\hat{B}\,\right\} \hat{\ell}_{ij}\;,
\label{deformedXX}
\end{equation}
where we have used the shorthand $\hat{A}=A(\hat{\mathbf{p}}^2)$, $\hat{A}'=\dfrac{dA}{d\mathbf{p}^2}(\hat{\mathbf{p}}^2)$, $\hat{B}=B(\hat{\mathbf{p}}^2)$, 
and $\hat{\ell}_{ij} = \left(\hat{x}_i\hat{p}_j-\hat{x}_j\hat{p}_i\right)/\hat{A}$. 
Here, $\hat{\ell}_{ij}$ is the angular momentum operator which generates rotations,
as can be seen from its commutation relations which are
\begin{eqnarray}
\dfrac{1}{i\hbar}[\,\hat{\ell}_{ij},\,\hat{x}_{k}\,]
& = & \delta_{ik}\hat{x}_j - \delta_{jk}\hat{x}_i \;,
\cr
\dfrac{1}{i\hbar}[\,\hat{\ell}_{ij},\,\hat{p}_{k}\,]
& = & \delta_{ik}\hat{p}_j - \delta_{jk}\hat{p}_i \;,
\cr
\dfrac{1}{i\hbar}[\,\hat{\ell}_{ij},\,\hat{\ell}_{mn}\,]
& = & \delta_{im}\hat{\ell}_{jn} - \delta_{in}\hat{\ell}_{jm}
    + \delta_{jn}\hat{\ell}_{im} - \delta_{jm}\hat{\ell}_{in}
\;.
\end{eqnarray}
Note that the non-commutativity of the components of position
can be interpreted as a reflection of the dynamic nature of space itself,
as would be expected in quantum gravity.

Various choices for the functions $A(\mathbf{p}^2)$ and $B(\mathbf{p}^2)$ 
have been considered in the literature.\cite{Snyder:1946qz,Maggiore:1993zu,Kempf:1994su,Scardigli:1999jh,Brau:1999uv,Brau:2006ca,Ali:2009zq,Das:2010zf,Ali:2011fa}
Here, we choose\footnote{%
Another popular choice in the literature has been 
\[
A(\mathbf{p}^2)\;=\; 1+\beta\mathbf{p}^2\;,\qquad
B(\mathbf{p}^2)\;=\; 2\beta\;,
\]
which leads to $[\hat{x}_i,\,\hat{x}_j\,]=O(\beta^2)$,
rendering this commutator negligible if only the leading order corrections in $\beta$ are considered.\cite{Brau:1999uv,Brau:2006ca}
}
\begin{equation}
A(\mathbf{p}^2)\;=\; 1-\beta\mathbf{p}^2\;,\qquad
B(\mathbf{p}^2)\;=\; 2\beta\;,
\label{OurAB}
\end{equation}
which leads to the algebra
\begin{eqnarray}
[\,\hat{p}_i,\,\hat{p}_j\,] & = & 0 \vphantom{\Big|}
\;,\cr
\dfrac{1}{i\hbar}[\,\hat{x}_i,\,\hat{p}_j\,]
& = & (1-\beta\hat{\mathbf{p}}^2)\delta_{ij}+2\beta\,\hat{p}_i\hat{p}_j
\;,\cr
\dfrac{1}{i\hbar}[\,\hat{x}_i,\,\hat{x}_j\,]
& = & 4\beta\hat{\ell}_{ij}\;.
\label{MLUR-CommutationRelations}
\end{eqnarray}
In 1D this reduces to the form we assumed in Eq.~(\ref{MLUR3}):
\begin{equation}
\dfrac{1}{i\hbar}[\,\hat{x},\,\hat{p}\,]
\;=\; (1-\beta\hat{p}^2)+2\beta\,\hat{p}^2
\;=\; 1+\beta\hat{p}^2
\;.
\end{equation}
For higher dimensions, the diagonal commutators read
\begin{equation}
\dfrac{1}{i\hbar}[\,\hat{x}_i,\,\hat{p}_i\,]
\;=\; (1+\beta\hat{p}_i^2) -\beta\sum_{j\neq i}\hat{p}_j^2\;,
\end{equation}
(no summation of repeated indices)
which implies 
\begin{equation}
\delta x_i\,\delta p_i \;\ge\;\dfrac{1}{2}
\Bigl|
\vev{[\,\hat{x}_i,\,\hat{p}_i\,]}
\Bigr|
\;=\; \dfrac{\hbar}{2}
\Bigl|
1 + \beta\vev{\hat{p}_i^2} - \beta\sum_{j\neq i}\vev{\hat{p}_j^2}
\Bigr|
\;.
\end{equation}
If $\vev{\hat{p}_j^2}=0$ for $j\neq i$, this in turn
implies $\delta x_i \sim \delta p_i$ for $\delta p_i\gg 1/\sqrt{\beta}$.
Despite this observation, summing over $i$ gives us
\begin{eqnarray}
\sum_{i=1}^{D}\delta x_i\,\delta p_i
& \ge & \dfrac{\hbar}{2}
\sum_{i=1}^{D}
\Bigl|
\vev{[\,\hat{x}_i,\,\hat{p}_i\,]}
\Bigr|
\cr
& \ge &  \dfrac{\hbar}{2}
\biggl|
\sum_{i=1}^{D}
\vev{[\,\hat{x}_i,\,\hat{p}_i\,]}
\biggr|
\;=\; \dfrac{\hbar}{2}
\biggl|
D - \beta(D-2)\vev{\hat{\mathbf{p}}^2}
\biggr|
\;,
\end{eqnarray}
which suggests that for $D\ge 3$, unlike the $D=1$ case, 
we should take $\beta<0$ for a minimal length to exist. 
For the borderline $D=2$ case, the lower bound 
on $\sum_{i=1}^{D}\delta x_i\,\delta p_i$ is independent of 
$\beta$ or $\vev{\hat{\mathbf{p}}^2}$, 
despite the deformation of the algebra, 
suggesting that $\beta$ can be assumed to be of either sign,
but the existence of the minimal length in this particular case is
obscure. 
We have nevertheless introduced a new length scale into the algebra
and continue to refer to $\hbar\sqrt{|\beta|}$ as the minimal length in this case also.

The operators which satisfy the algebra of Eq.~(\ref{MLUR-CommutationRelations})
can be represented by differential operators in 
$D$-dimensional $p$-space as
\begin{eqnarray}
\hat{p}_i & = & p_i \;, \vphantom{\Big|}\cr
\hat{x}_i & = & i\hbar
\left[ (1 - \beta \mathbf{p}^2)\frac{\partial}{\partial p_i}
       + 2\beta p_i p_j \frac{\partial}{\partial p_j}
       + \beta\left(D-\delta\right) \,p_i
\right] \;,\cr
\hat{\ell}_{ij} & = & -i\hbar\left(p_i\dfrac{\partial}{\partial p_j}-p_j\dfrac{\partial}{\partial p_i}\right)\;,
\end{eqnarray}
(repeated index is summed)
with the inner product of the $p$-space wave-functions given by
\begin{equation}
\langle f|g\rangle_\delta
\;=\; \int \dfrac{d^D\mathbf{p}}{(1+\beta\mathbf{p}^2)^\delta}\;f^*(\mathbf{p})\,g(\mathbf{p})\;.
\end{equation}
This form is required to render the $\hat{x}_i$ operators symmetric,
where the parameter $\delta$ appearing in the representation of $\hat{x}_i$ and
the integration measure is arbitrary.
By choosing $\delta=0$ we can simplify the integration measure at the expense of adding the extra term $\beta D p_i$ to the representation of $\hat{x}_i$,
which suggests that the singularity in the integration measure at
$\mathbf{p}^2=-1/\beta$ when $\beta$ is negative is removable,
and taking $\beta$ negative would not cause any intrinsic problems
in any number of dimensions as far as the 
differential operator representation is concerned.

Our choice of the functions $A(\mathbf{p}^2)$ and $B(\mathbf{p}^2)$ in Eq.~(\ref{OurAB})
has been motivated by the fact that it leads to a particularly simple form for the
commutator of the position operators, Eq.~(\ref{MLUR-CommutationRelations}),
allowing it to close on the angular momentum operator.
This suggests that the position operators of this algebra may be realizable as
angular momentum ($\beta >0$) or boost ($\beta<0$) operators in a higher dimension,
as already pointed out in the pioneering papers of Snyder\cite{Snyder:1946qz}
and Yang\cite{Yang:1947ud}.
Indeed, we show in the appendix that the $\hat{x}_i$ operators in $D$-dimensions can be realized as rotation/boost operators in $(D+1)$-dimensions, 
while the $\hat{p}_i$ operators can be realized as operators obtained by stereographic projection of a
$(D+1)$-dimensional hypersphere ($\beta>0$) or hyperboloid ($\beta<0$) down to $D$-dimensions,
$1/\sqrt{\beta}$ corresponding to the radius of the hypersphere.\cite{Yang:1947ud}

Taking the limit $\beta \to 0$, $\ell_{ij} \to \infty$ while keeping 
$4\beta\ell_{ij} \equiv \theta_{ij}$
fixed to constants leads to the algebra 
\begin{eqnarray}
[\,\hat{p}_i,\,\hat{p}_j\,] & = & 0 
\;,\vphantom{\Big|}\cr
[\,\hat{x}_i,\,\hat{p}_j\,]
& = & i\hbar\,\delta_{ij}
\;,\vphantom{\Big|}\cr
[\,\hat{x}_i,\,\hat{x}_j\,]
& = & i\hbar\,\theta_{ij}
\;.\vphantom{\Big|}
\end{eqnarray}
This limit can be understood geometrically as taking both the radius-squared of the hypersphere (hyperboloid) 
$1/\beta$ and the angular momenta $\ell_{ij}$ to infinity, while keeping their ratio fixed.
This is the algebra 
assumed in NCFT \cite{Douglas:2001ba,Szabo:2001kg,Polychronakos:2007df,Szabo:2009tn,Blaschke:2010kw,D'Ascanio:2016asa,Borowiec:2016zrc}
but with an important distinction.
When $\theta_{ij}$ are $c$-numbers, 
the $\hat{p}_i$ operator can be identified with the adjoint operator
\begin{equation}
\hat{p}_i \;=\; [\,-\omega_{ij}\hat{x}_j,\,*\,]\;,
\end{equation}
(repeated index is summed) where
\begin{equation}
\theta_{ik}\omega_{kj} \;=\;
\omega_{ik}\theta_{kj} \;=\; \delta_{ij}\;.
\end{equation}
Indeed
\begin{eqnarray}
[\,-\omega_{ik}\hat{x}_k,\,-\omega_{j\ell}\hat{x}_\ell\,] & = & -i\hbar\,\omega_{ij} 
\;,\vphantom{\Big|}\cr
[\,\hat{x}_i,\,-\omega_{jk}\hat{x}_k\,] & = & i\hbar\,\delta_{ij} 
\;,\vphantom{\Big|}
%\cr
%[\,\hat{x}_i,\,\hat{x}_j\,] & = & i\hbar\,\theta_{ij} \;.
\end{eqnarray}
and we find
\begin{eqnarray}
[\,\hat{p}_i,\,\hat{p}_j\,]\,*
& = & [\,-\omega_{ik}\hat{x}_k,\,[\,-\omega_{j\ell}\hat{x}_\ell,\,*\,]\,]-[\,-\omega_{j\ell}\hat{x}_\ell,\,[\,-\omega_{ik}\hat{x}_k,\,*\,]\,] 
\vphantom{\Big|}\cr
& = & [\,[\,-\omega_{ik}\hat{x}_k,\,-\omega_{j\ell}\hat{x}_\ell\,],\,*]
\vphantom{\Big|}\cr
& = & [\,-i\hbar\,\omega_{ij},\,*\,] \;=\; 0\;,
\vphantom{\Big|}\cr
%%%
[\,\hat{x}_i,\,\hat{p}_j\,]\,*
& = & \hat{x}_i[\,-\omega_{jk}\hat{x}_k,\,*\,] - [\,-\omega_{jk}\hat{x}_k,\,\hat{x}_i * \,]
\vphantom{\Big|}\cr
& = & [\,\hat{x}_i,\,-\omega_{jk}\hat{x}_k\,]\,*
\;=\; i\hbar\,\delta_{ij}*
\;,
\vphantom{\Big|}
\end{eqnarray}
for any operator $*$. 
Thus, the momentum operators $\hat{p}_i$ can be expressed in terms of the coordinate operators
$\hat{x}_i$, and this facilitates the introduction and understanding of gauge fields
in NCFT.
This identification cannot be performed in our algebra; our
`momentum space' stays distinct from `coordinate space,'
and though the NCFT algebra is a particular limiting case of ours, 
the question of gauge invariance takes on a different nature.
In particular, our formulation requires us to work in the full phase space 
instead of coordinate space.

%%%%%%%%%%%%%%%%%%%%%%%%%%%%%%%%%%%%%%%%%%%%%%%%%%%%%%%%%%%%%%%%%%%%%%%%%%%%%%%%%%%%%%%%%%%%%%%%%
\subsection{Introduction of the Gauge Field}

Let us now consider the introduction of the gauge field.
For simplicity we consider the case of the constant magnetic field.
We restrict our attention to the 2D case, which is the lowest dimension which would
allow us to introduce a uniform magnetic field, or its analog.
The introduction of a uniform magnetic field will also introduce a length scale, 
the magnetic length $\ell_M=\sqrt{\hbar/eB}$,\footnote{%
We set $c=1$.
} 
and its ratio to the minimal length $\ell_S=\hbar\sqrt{|\beta|}$ 
becomes an important parameter in what follows.
The algebra of the position, momentum, and angular momentum in 2D reads:
\begin{eqnarray}
[\,\hat{p}_i,\,\hat{p}_j\,] & = & 0\;,\vphantom{\Big|}\cr
[\,\hat{x}_i,\,\hat{p}_j\,] & = & 
i\hbar\left\{ 
(1-\beta\hat{\mathbf{p}}^2)\delta_{ij}+ 2\beta\hat{p}_i\hat{p}_j
\right\}
\;,\vphantom{\Big|}
\cr
[\,\hat{x}_i,\,\hat{x}_j\,] & = & 4i\hbar\beta\hat{\ell}_{ij} \;\equiv\; 4i\hbar\beta\epsilon_{ij}\hat{\ell}\;,\vphantom{\Big|}\cr
[\,\hat{\ell},\,\hat{x}_i\,] & = & i\hbar\epsilon_{ij}\hat{x}_j\;,\vphantom{\Big|}\cr
[\,\hat{\ell},\,\hat{p}_i\,] & = & i\hbar\epsilon_{ij}\hat{p}_j\;,\vphantom{\Big|}
\end{eqnarray}
where $i,j=1,2$.
As argued above, we allow $\beta$ to take on either sign.

%It is convenient to introduce the operators
%
%\begin{equation}
%\hat{x}_\pm \;\equiv\; \dfrac{\hat{x}_1\pm i\hat{x}_2}{\sqrt{2}}\;,\qquad
%\hat{p}_\pm \;\equiv\; \dfrac{\hat{p}_1\mp i\hat{p}_2}{\sqrt{2}}\;,
%\end{equation}
%
%in terms of which the above commutation relations take on a particularly simple form:
%
%\begin{eqnarray}
%[\,\hat{p}_+,\,\hat{p}_-\,] & = & 0\;,\vphantom{\Big|}\cr
%[\,\hat{x}_\pm,\,\hat{p}_\pm\,] & = & i\hbar \;,\vphantom{\Big|}\cr
%[\,\hat{x}_\mp,\,\hat{p}_\pm\,] & = & 2i\hbar\beta\,\hat{p}_\pm^2 \;,\vphantom{\Big|}\cr
%[\,\hat{x}_+,\,\hat{x}_-\,] & = & 4\hbar\beta\,\hat{\ell}\;,\vphantom{\Big|}\cr
%[\,\hat{\ell},\,\hat{x}_\pm\,] & = & \pm\hbar\,\hat{x}_\pm\;,\vphantom{\Big|}\cr
%[\,\hat{\ell},\,\hat{p}_\pm\,] & = & \mp\hbar\,\hat{p}_\pm\;.\vphantom{\Big|}
%\end{eqnarray}
%
%The free particle Hamiltonian can then be written as
%
%\begin{equation}
%\hat{H} 
%\;=\; \dfrac{\hat{\mathbf{p}}^2}{2m} 
%\;=\; \dfrac{\hat{p}_1^2+\hat{p}_2^2}{2m}
%\;=\; \dfrac{\hat{p}_+\hat{p}_- + \hat{p}_-\hat{p}_+}{2m}\;.
%\end{equation}
%

The covariant derivative is introduced as a shift of the momentum operator:
\begin{equation}
\hat{D}_i \;=\; \hat{p}_i + \hat{A}_i\;.
\end{equation}
For a gauge transformation on the state vector $\ket{\psi}$,
\begin{equation}
\ket{\psi} \;\rightarrow\; \ket{\psi'} \,=\, \hat{U}\ket{\psi}\;,
\end{equation}
the covariant derivative must also transform as
\begin{equation}
\hat{D}_i\ket{\psi} \;\rightarrow\; \hat{U}\hat{D}_i\ket{\psi} 
\;=\; 
\underbrace{\hat{U}\hat{D}_i\hat{U}^{-1}}_{\displaystyle \hat{D}_i'}
\underbrace{\hat{U}\ket{\psi}}_{\displaystyle \ket{\psi'}}
\;.
\end{equation}
Therefore,
\begin{eqnarray}
\hat{D} \;\rightarrow\; \hat{D}_i' 
& = & \hat{U}\hat{D}_i\hat{U}^{-1} \vphantom{\Big|}\cr
& = & \hat{U}(\hat{p}_i + \hat{A}_i)\hat{U}^{-1} \vphantom{\Big|}\cr
& = & \hat{p}_i + \hat{U}[\,\hat{p}_i,\,\hat{U}^{-1}\,] + \hat{U}\hat{A}_i\hat{U}^{-1} \vphantom{\Big|}\cr
& = & \hat{p}_i + \hat{A}'_i\;, \vphantom{\Big|}
\end{eqnarray}
that is
\begin{equation}
\hat{A}_i' \;=\; \hat{U}\hat{A}_i\hat{U}^{-1} + \hat{U}[\,\hat{p}_i,\,\hat{U}^{-1}\,] \;.
\end{equation}
The replacement
\begin{equation}
\hat{H}\,=\,\dfrac{\hat{\mathbf{p}}^2}{2m}
\quad\rightarrow\quad
\hat{H}' \,=\, \dfrac{\hat{\mathbf{D}}^2}{2m}
\end{equation}
then gives us a gauge covariant (not invariant) Hamiltonian: $\hat{H}\rightarrow\hat{H}'=\hat{U}\hat{H}\hat{U}^{-1}$.
The eigenvalues of $\hat{H}$ are invariant and physical since
it shares the same eigenvalues with $\hat{H}'$ :
\begin{equation}
\hat{H}\ket{\psi}\;=\;E\ket{\psi}
\quad\rightarrow\quad
\underbrace{(\hat{U}\hat{H}\hat{U}^{-1})}_{\displaystyle \hat{H}'}
\underbrace{\hat{U}\ket{\psi}}_{\displaystyle \ket{\psi'}}\;=\; E
\underbrace{\hat{U}\ket{\psi}}_{\displaystyle \ket{\psi'}}\;.
\end{equation}

In analogy with the $\beta=0$ case, we introduce the vector potential,
using 3D notation, as
\begin{equation}
\hat{\mathbf{A}} \;=\; \dfrac{e}{2}\;\mathbf{B}\times\hat{\mathbf{r}}
\;=\; \dfrac{eB}{2}
\begin{bmatrix} -\hat{x}_2 \\ \phantom{-}\hat{x}_1 \end{bmatrix}
\;,
\label{GaugeChoice}
\end{equation}
where $\mathbf{B}$ is the putative uniform magnetic field in the $z$-direction.
The magnetic field, defined in the usual fashion as a commutator of the covariant
derivative, becomes a gauge covariant operator
\begin{equation}
\hat{f}
\;=\; \dfrac{1}{i\hbar}\bigl[\,\hat{D}_1,\,\hat{D}_2\,\bigr]
\;=\; eB + (eB)^2\beta\hat{\ell}\;,
\label{fhatdef}
\end{equation}
which transforms as $\hat{f}\rightarrow\hat{f}'=\hat{U}\hat{f}\hat{U}^{-1}$.
We define the magnetic field strength as its trace:
\begin{equation}
f 
\;\equiv\; \mathrm{Tr}\hat{f} 
\;=\; \mathrm{Tr}\left[eB + (eB)^2\beta\hat{\ell}\,\right]
\;=\; eB\;,
\end{equation}
where the trace is over the Hilbert space, and we have assumed
\begin{equation}
\mathrm{Tr}\,\hat{1}\;=\;1\;,\qquad
\mathrm{Tr}\,\hat{\ell}\;=\;0\;.
\end{equation}
So the magnetic field strength is the same as the $\beta=0$ case.

Note that the vector potential of Eq.~(\ref{GaugeChoice}) satisfies the Coulomb gauge condition
\begin{eqnarray}
\sum_{i=1}^{2}[\,\hat{p}_i,\,\hat{A}_i\,] 
& = & \dfrac{eB}{2}
\Bigl(
 [\,\hat{p}_1,\,-\hat{x}_2\,]
+[\,\hat{p}_2,\, \hat{x}_1\,]
\Bigr) 
\cr
%& = & \dfrac{eB}{2}
%\Bigl(
% [\,\hat{x}_2,\,\hat{p}_1\,]
%-[\,\hat{x}_1,\,\hat{p}_2\,]
%\Bigr) 
%\cr
& = & i\hbar eB\beta
\Bigl(
 \hat{p}_2\hat{p}_1
-\hat{p}_1\hat{p}_2
\Bigr)
\;=\; 0\;,\vphantom{\bigg|}
\end{eqnarray}
so $\hat{\mathbf{p}}\cdot\hat{\mathbf{A}} = \hat{\mathbf{A}}\cdot\hat{\mathbf{p}}$.
We find
\begin{eqnarray}
\hat{H}'
& = & \dfrac{1}{2m}\left(\hat{\mathbf{p}}+\hat{\mathbf{A}}\right)^2
\cr
& = & \dfrac{1}{2m}
\left(
\hat{\mathbf{p}}^2 +
2\hat{\mathbf{A}}\cdot\hat{\mathbf{p}} + \hat{\mathbf{A}}^2
\right)
\cr
%& = & \dfrac{1}{2m}
%\left\{
%\hat{\mathbf{p}}^2 +
%e\left(\mathbf{B}\times\hat{\mathbf{r}}\right)\cdot\hat{\mathbf{p}}
%+ \dfrac{e^2}{4}\left(\mathbf{B}\times\hat{\mathbf{r}}\right)^2
%\right\}
%\cr
%& = & \dfrac{1}{2m}
%\left\{
%\hat{\mathbf{p}}^2 +
%e\left(\hat{\mathbf{r}}\times\hat{\mathbf{p}}\right)\cdot\mathbf{B}
%+ \dfrac{(eB)^2}{4}\hat{\mathbf{r}}^2
%\right\}
%\cr
& = & \dfrac{\hat{\mathbf{p}}^2}{2m}
+ \dfrac{eB}{2m}(\hat{x}_1\hat{p}_2-\hat{x}_2\hat{p}_1)
+ \dfrac{e^2}{8m}\,\hat{\mathbf{r}}^2 B^2
\cr
& = &
\dfrac{\hat{\mathbf{p}}^2}{2m}
+ \dfrac{eB}{2m}\,(1-\beta\hat{\mathbf{p}}^2)\,\hat{\ell}
+ \dfrac{e^2}{8m}\,\hat{\mathbf{r}}^2 B^2
\cr
& = &
\dfrac{\hat{\mathbf{p}}^2}{2m}
+ \hat{\mu}B - \dfrac{1}{2}\hat{\alpha}B^2
\;,
\end{eqnarray}
where we have identified the magnetic dipole moment and the magnetic polarizability with
\begin{equation}
\hat{\mu} \;=\; \dfrac{e}{2m}\,(1-\beta\hat{\mathbf{p}}^2)\,\hat{\ell}\;,\qquad
\hat{\alpha} \;=\; -\dfrac{e^2}{4m}\,\hat{\mathbf{r}}^2 \;.
\label{mualpha}
\end{equation}
The dependence of the magnetic dipole moment on $\beta\hat{\mathbf{p}}^2$ is intriguing.
It implies that, for $\beta>0$, the magnetic dipole moment vanishes
when $\vev{\hat{\mathbf{p}}^2}=1/\beta$ and can even point in the opposite direction from the angular momentum when $\vev{\hat{\mathbf{p}}^2}>1/\beta$.
(Note that $\hat{\mathbf{p}}^2$ and $\hat{\ell}$ commute so they can be simultaneously diagonalized.)

Now, let us perform a gauge transformation.
Consider the transformation
\begin{equation}
\hat{U} \;=\; e^{-i\epsilon(\hat{x}_1\hat{x}_2+\hat{x}_2\hat{x}_1)/2\hbar}\;.
\label{GaugeTransformation1}
\end{equation}
To leading order in $\epsilon$ we find
\begin{eqnarray}
\hat{U}\hat{p}_1\hat{U}^{-1}
& = & \hat{p}_1   -\dfrac{i\epsilon}{2\hbar}[\,\hat{x}_1\hat{x}_2+\hat{x}_2\hat{x}_1,\,\hat{p}_1\,] + \cdots  \vphantom{\bigg|} \cr
%& = & \hat{p}_1 + \dfrac{\epsilon}{2}
%\left\{
% \hat{x}_2(1+\beta\hat{p}_1^2-\beta\hat{p}_2^2)
%+(1+\beta\hat{p}_1^2-\beta\hat{p}_2^2)\hat{x}_2
%\right\}
%+ \epsilon\beta
%\left(
% \hat{x}_1\hat{p}_2\hat{p}_1
%+\hat{p}_1\hat{p}_2\hat{x}_1
%\right)
%+ \cdots
%\cr
& = & \hat{p}_1  + \epsilon\hat{x}_2 
+ \dfrac{\epsilon\beta}{2}\left\{
2(\hat{x}_1\hat{p}_1\hat{p}_2+\hat{p}_1\hat{p}_2\hat{x}_1)
+\hat{x}_2\left(\hat{p}_1^2-\hat{p}_2^2\right)
+\left(\hat{p}_1^2-\hat{p}_2^2\right)\hat{x}_2
\right\}
+ \cdots \;,\vphantom{\bigg|} 
\cr
%& = & \hat{p}_1 + \epsilon\hat{x}_2
%+ \dfrac{\epsilon\beta}{2}\left\{
% 2(\hat{x}_1\hat{p}_2+\hat{x}_2\hat{p}_1)\hat{p}_1
%+2\hat{p}_1(\hat{p}_2\hat{x}_1+\hat{p}_1\hat{x}_2)
%-(\hat{x}_2\hat{\mathbf{p}}^2 +\hat{\mathbf{p}}^2\hat{x}_2)
%\right\}
%+ \cdots
%\cr
%%%
\hat{U}\hat{A}_1\hat{U}^{-1}
& = & -\dfrac{eB}{2}\hat{U}\hat{x}_2\hat{U}^{-1}
\vphantom{\bigg|}  \cr
%& = & -\dfrac{eB}{2}\left[
%\hat{x}_2 - \dfrac{i\epsilon}{2\hbar}
%[\,\hat{x}_1\hat{x}_2+\hat{x}_2\hat{x}_1,\,\hat{x}_2\,] + \cdots
%\right]
%\cr
& = & -\dfrac{eB}{2}\left[
\hat{x}_2 - 2\epsilon\beta\left( \hat{\ell}\hat{x}_2 + \hat{x}_2\hat{\ell}\right) + \cdots
\right]
\;,\vphantom{\bigg|} 
\end{eqnarray}
with similar expressions for $\hat{U}\hat{p}_2\hat{U}^{-1}$ 
and $\hat{U}\hat{A}_2\hat{U}^{-1}$.
When $\beta=0$, we have
\begin{eqnarray}
\hat{U}(\hat{p}_1+\hat{A}_1)\hat{U}^{-1}
& \quad\xrightarrow{\beta=0}\quad & 
\hat{p}_1 - \left(\dfrac{eB}{2}-\epsilon\right)\hat{x}_2 
\;, \vphantom{\Bigg|}
\cr
\hat{U}(\hat{p}_2+\hat{A}_2)\hat{U}^{-1}
& \quad\xrightarrow{\beta=0}\quad & 
\hat{p}_2 + \left(\dfrac{eB}{2}+\epsilon\right)\hat{x}_1
\;, \vphantom{\Bigg|}
\end{eqnarray}
with $\epsilon=\pm\dfrac{eB}{2}$ leading to Landau gauge.
For the $\beta\neq 0$ case, the components of the gauge transformed vector potential 
$\hat{A}_i'$ ($i=1,2$)
become infinite series involving both position and momentum operators.
They are intrinsically non-local operators.

Though the gauge transformation operator, Eq.~(\ref{GaugeTransformation1}),
depends only on the position operators, 
the transformation itself is highly non-local.
This suggests a generalization of gauge transformations 
to those that are non-local (in $x$-space) even in the $\beta=0$ limit, such as:
\begin{equation}
\hat{U} \;=\; e^{-i\epsilon\hat{p}_1\hat{p}_2/\hbar}\;,\quad
\hat{U} \;=\; e^{-i\epsilon(\hat{x}_1\hat{p}_1+\hat{x}_2\hat{p}_2)/\hbar}\;,\quad
\mbox{etc.}
\end{equation}
The rotation operator $\hat{U}=\exp[-i\theta\hat{\ell}/\hbar]$ will then be just a particular 
``gauge transformation'' which leaves the Hamiltonian invariant.
This type of generalization may be required of us in the presence of a minimal length,
and it also merges nicely with the ideas of 
UV/IR correspondence\cite{Green:1987sp,Green:1987mn,Polchinski:1998rq,Polchinski:1998rr,Becker:2007zj,Susskind:1998dq,Peet:1998wn}, 
Born reciprocity\cite{Born:1935,Born:1949}, etc.
which are expected in that context.

The traceless term in $\hat{f}$, Eq.~(\ref{fhatdef}), is dependent on the ratio of the minimal and magnetic lengths:
\begin{equation}
\ell_S \;=\; \hbar\sqrt{|\beta|}\;,\qquad \ell_M \;\equiv\; \sqrt{\dfrac{\hbar}{eB}}\;,\end{equation}
and we can write
\begin{equation}
\hat{f} \;=\; eB \left[ 1 \pm \left(\dfrac{\ell_S}{\ell_M}\right)^2 \hat{L} \right]\;,\qquad
\hat{L}\;=\;\dfrac{\hat{\ell}}{\hbar}\;,
\end{equation}
where the $\pm$ refers to the sign of $\beta$.
The Hamiltonian of a particle of mass $m$ coupled to the gauge field is 
\begin{equation}
\hat{H}' 
\;=\; \dfrac{\hat{D}_1^2 + \hat{D}_2^2}{2m}
\;=\; \dfrac{\hat{D}_+\hat{D}_- + \hat{D}_-\hat{D}_+}{2m} 
\;,
\end{equation}
where
%
%\begin{eqnarray}
%\hat{D}_\pm & \equiv & \dfrac{\hat{D}_1 \mp i\hat{D}_2}{\sqrt{2}} \;,\cr
%\hat{D}_1 & = & \hat{p}_1 + \dfrac{eB}{2}\hat{x}_2\;,\cr
%\hat{D}_2 & = & \hat{p}_2 - \dfrac{eB}{2}\hat{x}_1\;,
%\end{eqnarray}
%
\begin{equation}
\hat{D}_\pm \;\equiv\; \dfrac{\hat{D}_1 \mp i\hat{D}_2}{\sqrt{2}} \;,\quad
\hat{D}_1 \;=\; \hat{p}_1 + \dfrac{eB}{2}\hat{x}_2\;,\quad
\hat{D}_2 \;=\; \hat{p}_2 - \dfrac{eB}{2}\hat{x}_1\;,
\end{equation}
and
\begin{equation}
\dfrac{1}{i\hbar}[\,\hat{D}_1,\,\hat{D}_2\,]
\;=\; \dfrac{1}{\hbar}[\,\hat{D}_-,\,\hat{D}_+\,]
\;=\; \hat{f}
\;=\; eB\left[1\pm\left(\dfrac{\ell_S}{\ell_M}\right)^2\hat{L}\right]
\;.
\end{equation}
Consider a subspace of the Hilbert space spanned by the eigenvectors of the angular momentum operator $\hat{L}$ with eigenvalue $L$.  
In that space, the field strength operator $\hat{f}$ can be replaced by the number
\begin{equation}
\bvev{\hat{f}}
\;=\; 
f_L \;\equiv\; 
eB\Bigl[\,1+(eB\hbar\beta)L\,\Bigr] \;=\;
eB\left[1\pm\left(\dfrac{\ell_S}{\ell_M}\right)^2 L\right]\;.
\end{equation}
Note that $f_L$ can be negative depending on the value and sign of $\beta L$.
Let us define
\begin{equation}
\hat{a}_L^{\phantom{\dagger}} \;\equiv\; \dfrac{\hat{D}_-}{\sqrt{\hbar f_L}} \;,\qquad
\hat{a}_L^\dagger             \;\equiv\; \dfrac{\hat{D}_+}{\sqrt{\hbar f_L}} \;,%\qquad
%[\,\hat{a}_L^{\phantom{\dagger}},\,\hat{a}_L^\dagger\,] \;=\; 1\;,
\end{equation}
when $f_L>0$, and
\begin{equation}
\hat{a}_L^{\phantom{\dagger}} \;\equiv\; \dfrac{\hat{D}_+}{\sqrt{\hbar |f_L|}} \;,\qquad
\hat{a}_L^\dagger             \;\equiv\; \dfrac{\hat{D}_-}{\sqrt{\hbar |f_L|}} \;,%\qquad
%[\,\hat{a}_L^{\phantom{\dagger}},\,\hat{a}_L^\dagger\,] \;=\; 1\;,
\end{equation}
when $f_L<0$. 
In both cases we have
\begin{equation}
[\,\hat{a}_L^{\phantom{\dagger}},\,\hat{a}_L^\dagger\,] \;=\; 1\;,
\end{equation}
and the Hamiltonian restricted to the $L$-subspace, $\hat{H}'_L$, can be written as
\begin{equation}
\hat{H}'_L 
\;=\; \hbar\omega_L\left(\dfrac{\hat{a}_L^{\phantom{\dagger}}\hat{a}_L^\dagger + \hat{a}_L^\dagger\hat{a}_L^{\phantom{\dagger}}}{2}\right)
\;=\; \hbar\omega_L\left(\hat{a}_L^\dagger\hat{a}_L^{\phantom{\dagger}} + \dfrac{1}{2}\right)\;,
\end{equation}
where
\begin{equation}
\omega_L \;=\; \dfrac{|f_L|}{m} 
\;=\; \omega_c\left|1\pm\left(\dfrac{\ell_S}{\ell_M}\right)^2 L\right|  \;,\qquad
\omega_c \;=\; \dfrac{eB}{m}\;.
\end{equation}
Thus, the energy eigenvalues are 
\begin{equation}
E_{L,n} \;=\; \hbar\omega_L\left(n+\dfrac{1}{2}\right)\;,\qquad
n\;=\;0,1,2,\cdots
\end{equation}
They are evenly spaced in each $L$-subspace with the separation depending on 
the ratio of the minimal to magnetic lengths $\ell_S/\ell_M$, the angular momentum $L$,
and the sign of $\beta$.
And interesting case is when $f_L=0$. 
In that case, $\hat{D}_1$ and $\hat{D}_2$ commute, and the
Hamiltonian $\hat{H}'_L$ for that value of $L$ is the free particle Hamiltonian.

%%%%%%%%%%%%%%%%%%%%%%%%%%%%%%%%%%%%%%%%%%%%%%%%%%%%%%%%%%%%%%%%%%%%%%%%%%%%%%%%%%%%%%%%%%%%%%%%%
\subsection{Introduction of the Gauge Field -- Alternative Approach}

The algebra involving $\beta$ is covariant under changes in parametrization in $p$-space, so to a large extent, we may choose the functions $A({\bf{p}}^2)$ to be anything we like.   The situation is different however if we choose to change the parametrization of both ${\bf p}$ and $x$.   Indeed, for some singular cases, we can recover the Heisenberg algebra for restricted domains over which the operators are defined.

We begin by introducing:
\begin{eqnarray}
\hat{\Pi}_\pm
%& \equiv & \dfrac{1}{\sqrt{2}}
%\left(\dfrac{\hat{p}_1 \mp i\hat{p}_2}{1+\beta\hat{\mathbf{p}}^2}\right)
%\;,
%\cr
& \equiv & \dfrac{1}{1+\beta\hat{\mathbf{p}}^2}
\left(\dfrac{\hat{p}_1 \mp i\hat{p}_2}{\sqrt{2}}\right)
\;,\vphantom{\Bigg|}
\cr
%%%
\hat{\Xi}_\pm 
& \equiv & \dfrac{1}{2}
\left\{
\left(\dfrac{1+\beta\hat{\mathbf{p}}^2}{1-\beta\hat{\mathbf{p}}^2}\right)
\left(\dfrac{\hat{x}_1 \pm i\hat{x}_2}{\sqrt{2}}\right)
+
\left(\dfrac{\hat{x}_1 \pm i\hat{x}_2}{\sqrt{2}}\right)
\left(\dfrac{1+\beta\hat{\mathbf{p}}^2}{1-\beta\hat{\mathbf{p}}^2}\right)
\right\}
\;.
\end{eqnarray}
It is straightforward to show that these operators satisfy the canonical commutation relations
\begin{eqnarray}
[\,\hat{\Xi}_+,\,\hat{\Xi}_-\,]
& = & [\,\hat{\Pi}_+,\,\hat{\Pi}_-\,]
\;=\; 0 \;,\vphantom{\Big|}\cr
[\,\hat{\Xi}_\pm,\,\hat{\Pi}_\pm\,] & = & i\hbar \;,\vphantom{\Big|}\cr
[\,\hat{\Xi}_\pm,\,\hat{\Pi}_\mp\,] & = & 0 \;.\vphantom{\Big|}
\end{eqnarray}
Note that there is no $\beta$ dependence in this algebra, which is identical to the conventional Heisenberg algebra, with $\beta=0$.    From the perspective of the approach discussed in the appendix, these operators arise by projection of a paraboloid, instead of a sphere ($\beta>0$, or hyperboloid ($\beta <0 $).   
In this context, the domains of the operators are restricted to $\mathbf{p}^2 < 1/\beta$.

The angular momentum operator can be expressed using these operators as
\begin{equation}
\hat{\ell} \;=\; i\,(\,\hat{\Xi}_+\hat{\Pi}_+ -\hat{\Xi}_-\hat{\Pi}_-) \;.
\end{equation}
Define the deformed free-particle Hamiltonian as
\begin{equation}
\hat{H}_\beta 
\;=\; 
\dfrac{1}{2m}\left(\hat{\Pi}_+\hat{\Pi}_- + \hat{\Pi}_-\hat{\Pi}_+\right)
\;=\;
\dfrac{1}{2m}
\dfrac{\hat{p}_1^2 + \hat{p}_2^2}{(1+\beta\hat{\mathbf{p}}^2)^2}
\;,
\end{equation}
which commutes with the $\hat{p}_i$'s, as is required for the particle to be `free,'
and converges to the usual Hamiltonian in the limit $\beta\rightarrow 0$.

The vector potential is introduced as 
\begin{equation}
\hat{A}_+ \;=\; -i\xi_1 eB\,\hat{\Xi}_-\;,\qquad
\hat{A}_- \;=\; +i\xi_2 eB\,\hat{\Xi}_+\;,
\end{equation}
where $\xi_1+\xi_2=1$, and the covariant derivative as
\begin{eqnarray}
\hat{D}_+ & = & \hat{\Pi}_+ + \hat{A}_+ \;=\; \hat{\Pi}_+ -i\xi_1 eB\,\hat{\Xi}_-
\;,\vphantom{\Big|}\cr
\hat{D}_- & = & \hat{\Pi}_- + \hat{A}_- \;=\; \hat{\Pi}_- +i\xi_2 eB\,\hat{\Xi}_+
\;.
\label{Dpmdef}
\end{eqnarray}
Then
\begin{equation}
\hat{f} \;=\;
\dfrac{1}{\hbar}[\,\hat{D}_-,\,\hat{D}_+\,]
\;=\; (\xi_1+\xi_2)\,eB
\;=\; eB
\;.
\end{equation}
Thus, a more generic gauge can be written down compactly in this approach,
and the field strength is gauge invariant without the taking of any trace.
Gauge transformation from one gauge to another is effected by the operator
\begin{equation}
\hat{U} \;=\; e^{\epsilon(\hat{\Xi}_+^2 - \hat{\Xi}_-^2)/2\hbar}\;,
\end{equation}
which shifts the gauge parameters: 
$\xi_1\rightarrow\xi_1-\epsilon$,
$\xi_2\rightarrow\xi_2+\epsilon$.
For the symmetric $\xi_1=\xi_2=\frac{1}{2}$ case, 
the Hamiltonian in the presence of the gauge field can be written as
\begin{eqnarray}
\hat{H}'_\beta 
& = & \dfrac{1}{2m}(\,\hat{D}_+\hat{D}_- + \hat{D}_-\hat{D}_+) 
\vphantom{\bigg|}\cr
& = & \dfrac{(\,\hat{\Pi}_+\hat{\Pi}_- + \hat{\Pi}_-\hat{\Pi}_+)}{2m}
+ \dfrac{eB}{2m}\,\hat{\ell} 
+ \dfrac{e^2 B^2}{8m}(\,\hat{\Xi}_+\hat{\Xi}_- + \hat{\Xi}_-\hat{\Xi}_+)
\;.
\end{eqnarray}
Since the $\hat{\Xi}$'s and $\hat{\Pi}$'s are canonically commuting,
this Hamiltonian is exactly the same as the $\beta=0$ case.  But it is defined over a restricted domain where $\beta{\bf p}^2 < 1$ for $\beta >0$. and  so the corresponding eigenvalues are expected to be different.  The presence of the minimal length is encoded in this restriction.

Nonetheless, this Hamiltonian will exhibit the infinite degeneracy present when $\beta =0$.   
Consider the following operators:
\begin{eqnarray}
\hat{Q}_+ & \equiv & \hat{\Pi}_+ + i \xi_2 eB\,\hat{\Xi}_- \;,\vphantom{\Big|}\cr
\hat{Q}_- & \equiv & \hat{\Pi}_- - i \xi_1 eB\,\hat{\Xi}_+ \;.\vphantom{\Big|}
\end{eqnarray}
Note the different coefficients of the $\hat{\Xi}$'s from Eq.~(\ref{Dpmdef}).
These operators satisfy the commutation relations
\begin{eqnarray}
\dfrac{1}{\hbar}[\hat{Q}_-, \hat{Q}_+] & = & (\xi_1 + \xi_2) eB \;=\; eB 
\;,\vphantom{\bigg|}\cr
%%%
[\,\hat{Q}_\pm,\,\hat{D}_\pm\,] & = &
[\,\hat{Q}_\pm,\,\hat{D}_\mp\,] \;=\; 0\;.\vphantom{\Big|}
\end{eqnarray}
That is, $\hat{Q}_\pm$ commute with the Hamiltonian but not with each other. 
The result is the infinite degeneracy already present in the normal case when $\beta=0$.  

We are yet to understand the full implication of this:
the extent to which the minimal length has been `gauged away,' so to speak, in this approach, or
whether the singularities in $p$-space are physical and lead to observable consequences at the
Planck scale.  
It is also unclear whether the existence of the canonically commuting operators is restricted to 2D.
These issues will be explored in an upcoming paper.\cite{NextOne}

%%%%%%%%%%%%%%%%%%%%%%%%%%%%%%%%%%%%%%%%%%%%%%%%%%%%%%%%%%%%%%%%%%%%%%%%%%%%%%%%%%%%%%%%%%%%%%%%%
\section{Summary and Discussion}

Note that our discussion generalizes the notion of gauge invariance found in the context of non-commutative field theory (NCFT) \cite{Douglas:2001ba,Szabo:2001kg,Polychronakos:2007df,Szabo:2009tn,Blaschke:2010kw,D'Ascanio:2016asa,Borowiec:2016zrc}. Given the geometric formulation 
presented in the appendix, it should be possible to deform the usual
discussion from NCFT based on the Moyal product by realizing gauge transformations in terms of
the isometries of the curved momentum space, associated with either a spherical or hyperbolic geometry.
Here we give a scenario for a more general formulation that would include the geometry of the momentum space.

The first lesson of our discussion is that, in the presence of a minimal length, we should formulate quantum field theory (and thus gauge field theory) in momentum space.
Then we should take into account the curved geometry of the momentum space, the curvature being directly tied to the minimal length parameter $\beta$.
Thus the classical action for such a momentum space field theory (for a scalar field $\phi$ in $3+1$ dimensions) would read (see Ref.~\citenum{Matsuo:2005fb})
\begin{equation}
S_m \;=\; 
\int dt\;\frac{d^3 \mathbf{p}}{(1+\beta \mathbf{p}^2)^3} 
\left[\,
\pi(\mathbf{p}, t)\,\dot{\phi}(\mathbf{p}, t) - H\left( \phi(\mathbf{p}, t), \pi(\mathbf{p}, t) \right)
\,\right]
\end{equation}
where $\pi (\mathbf{p}, t)$ is the conjugate momentum and $H$ the Hamiltonian density.
The gauge field would appear in momentum space as a generalized 1-form:
\begin{equation}
\pi(\mathbf{p}, t) \;\to\; 
\pi(\mathbf{p}, t) + A(\mathbf{p}, t) \;\equiv\; D(\mathbf{p}, t) \;.
\end{equation}
This suggests that the Yang-Mills theory would have to be formulated in momentum space, the curvature being given by the commutators of the above $D(\mathbf{p}, t) $.
Finally, the quantum theory would be defined in terms of the phase space path integral over the curved momentum space 
\begin{equation}
\int \mathcal{D}\phi(\mathbf{p}, t)\, \mathcal{D}\pi(\mathbf{p}, t)\,e^{\frac{i}{\hbar} S_m}
\end{equation}
where the measure in the path integral includes the geometry of the curved momentum space 
\begin{equation}
\mathcal{D}\phi(\mathbf{p}, t)\,
\mathcal{D}\pi (\mathbf{p}, t) 
\;\equiv\; 
\prod_i \prod_p \frac{d^3 \mathbf{p}}{(1+\beta \mathbf{p}^2)^3}
\, d\phi(\mathbf{p},t_i)
\, d\pi (\mathbf{p},t_i) 
\;.
\end{equation}
In principle, such a path integral defines the required generalization of the Moyal product in our more general case.

The discussion of this note was entirely non-relativistic. Obviously the next step has to involve the issue of relativistic covariance.
The usual NCFT based on the Moyal product \cite{Douglas:2001ba,Szabo:2001kg,Szabo:2009tn,Blaschke:2010kw,D'Ascanio:2016asa,Borowiec:2016zrc} has some outstanding features, such as the essential non-Abelian nature of even
a $U(1)$ formulation, as well as the existence of a two-scale renormalization group that was required for the proper definition of the continuum limit \cite{Grosse:2004yu}. As we have mentioned, this canonical NCFT can be understood as a particular case of the algebraic structure that involves the minimal length $\beta$ and the angular momentum $L_{ij}$ found in our discussion.
Note that in our case the non-Abelian nature is also present quite generically, but the existence of a two-scale renormalization is not clear.
For that we would have to make a more precise sense of the above path integral with a Wilsonian cut-off and follow the Wilson-Polchinski-like discussion of \cite{Grosse:2004yu}.
We note that the existence of a curved and dynamical momentum space as well as the two-scale renormalization is also expected on more fundamental grounds in the recent discussion
of quantum gravity in the context of metastring theory  \cite{Freidel:2013zga,Freidel:2014qna,Freidel:2015pka,Freidel:2015uug}, which might be underlying the more effective analysis presented in this note.
Nevertheless, the considerations of this note point to an interesting interplay of spatial non-commutativity, curved momentum space and gauge invariance in the context of
the minimal length physics.

In conclusion, in this paper, we have attempted to shed new light on the issue of the minimal
length. Starting from the well-understood concept of the magnetic
length, which also serves as the basis for the minimal length in
non-commutative field theory (NCFT), we developed a new realization of the
spatial minimal length without and with the presence of an external 
magnetic field. The intuition here is that a uniform magnetic field through an infinite plane can be mapped to a magnetic field of a
magnetic monopole through a sphere of a fixed radius. The sphere and the infinite plane are related through a stereographic projection.
Thus, in order to realize the minimal spatial length, we start with the symmetries of a sphere and then induce the algebra of position and momentum operators with a minimal length, by implementing stereographic projection from that sphere onto an infinite plane.
The same procedure can be realized by a stereographic projection from a hyperboloid onto an infinite plane.
(This approach can in principle be covariantized, even though we do not discuss that topic in the present paper.)
This procedure provides us with a new view on the origin of the minimal length and points to a formulation of physics that relies on a curved (spherical or hyperbolical) momentum-like space.
By gauging this construction in the presence of a uniform magnetic field, we find an interesting interplay between the minimal spatial and magnetic lengths. 
The overall approach gives us a possible new handle on the experimental searches for the minimal length, which have become more topical and much more realistic given the recent remarkable breakthroughs achieved by the LIGO collaboration,\cite{Abbott:2016blz} the gravitational interferometry being the natural experimental playground in this arena.

%%%%%%%%%%%%%%%%%%%%%%%%%%%%%%%%%%%%%%%%%%%%%%%%%%%%%%%%%%%%%%%%%%%%%%%%%%%%%%%%%%%%%%%%%%%%%%%%%
\section*{Acknowledgments}

The work of DM is supported in part by
the U.S. Department of Energy, grant DE-FG02-13ER41917, task A.

%%%%%%%%%%%%%%%%%%%%%%%%%%%%%%%%%%%%%%%%%%%%%%%%%%%%%%%%%%%%%%%%%%%%%%%%%%%%%%%%%%%%%%%%%%%%%%%%%
%%%%%%%%%%%%%%%%%%%%%%%%%%%%%%%%%%%%%%%%%%%%%%%%%%%%%%%%%%%%%%%%%%%%%%%%%%%%%%%%%%%%%%%%%%%%%%%%%
%\newpage
\appendix{Projection from a Higher-Dimensional Momentum Space}

In order to simply our notation, let us first introduce dimensionless operators by
\begin{equation}
\hat{X}_i \;=\; \dfrac{\hat{x}_i}{\hbar\sqrt{|\beta|}}\;,\qquad
\hat{P}_i \;=\; \sqrt{|\beta|}\,\hat{p}_i\;,\qquad
\hat{L}_{ij} \;=\; \dfrac{\hat{\ell}_{ij}}{\hbar}\;.
\end{equation}
Here, we allow for the possibility that $\beta$ is negative.
The algebra in terms of these dimensionless operators is
\begin{eqnarray}
[\,\hat{P}_i,\,\hat{P}_j\,] & = & 0 \vphantom{\Big|}
\;,\cr
\dfrac{1}{i}[\,\hat{X}_i,\,\hat{P}_j\,]
& = & (1\mp\hat{\mathbf{P}}^2)\delta_{ij}\pm 2\hat{P}_i\hat{P}_j
\;,\cr
\dfrac{1}{i}[\,\hat{X}_i,\,\hat{X}_j\,]
& = & \pm 4\hat{L}_{ij}\;,
\end{eqnarray}
where the upper(lower) sign corresponds to positive(negative) $\beta$.

\begin{figure}[t]
\begin{center}
\subfigure[$\beta>0$]{\includegraphics[width=5cm]{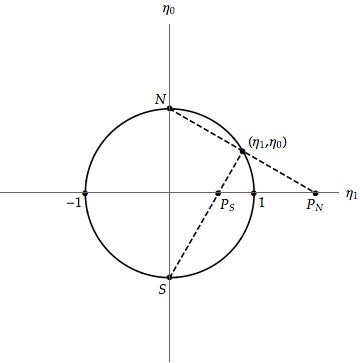}}\hspace{1cm}
\subfigure[$\beta<0$]{\includegraphics[width=5cm]{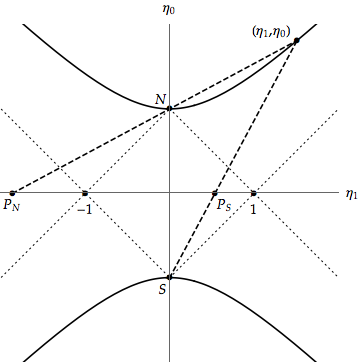}}
\caption{Stereographic projection from a $(D+1)$ dimensional (a) sphere, and (b) hyperboloid
down to $D$ dimensions, shown for the case $D=1$.
The north pole projection maps the northern hemisphere (or the northern branch of the hyperboloid) to the outside of the unit circle (sphere), and the southern hemisphere (southern branch of the hyperboloid) to the inside of the unit circle (sphere).
For the south pole projection it is the other way around.
A switch between the north and south pole projections corresponds to an inversion with respect to the unit circle (sphere) in $D$ dimensions.
}
\label{Projection}
\end{center}
\end{figure}

First, consider a unit $D$-dimensional sphere embedded in $(D+1)$-dimensional momentum space:
\begin{equation}
\hat{\eta}_0^2 + (\hat{\eta}_1^2+\hat{\eta}_2^2 + \cdots + \hat{\eta}_D^2) \;=\; 1\;,
\end{equation}
where we assume
\begin{eqnarray}
[\,\hat{\eta}_\alpha,\,\hat{\eta}_\beta\,] & = & 0\;,\vphantom{\Big|}\cr
[\,\hat{L}_{\alpha\beta},\,\hat{\eta}_{\gamma}\,] & = & 
i(\delta_{\alpha\gamma}\hat{\eta}_\beta - \delta_{\beta\gamma}\hat{\eta}_\alpha) \;.\vphantom{\Big|}
\label{etaCommutationRelations}
\end{eqnarray}
Note also
\begin{eqnarray}
[\,\hat{L}_{0i},\,\hat{L}_{0j}\,] & = & i\hat{L}_{ij}\;,\vphantom{\Big|}\cr
[\,\hat{L}_{ij},\,\hat{L}_{0k}\,] & = & i(\delta_{ik}\hat{L}_{0j}-\delta_{jk}\hat{L}_{0i})\;.\vphantom{\Big|}
\end{eqnarray}
Stereographic projection from the North(South) pole of the sphere to the
$\eta_0=0$ hyperplane yields
\begin{equation}
\hat{P}_{i} \;=\; \dfrac{\hat{\eta}_i}{1\mp\hat{\eta}_0}\;,\qquad
\hat{\mathbf{P}}^2 
\;=\; \sum_{i=1}^{D}\hat{P}_{i}^2
\;=\; \dfrac{1\pm\hat{\eta}_0}{1\mp\hat{\eta}_0}
\;,
\label{Pdefs}
\end{equation}
the upper(lower) sign corresponding to the North(South)-pole projection.
See Fig.~\ref{Projection}(a).
It is straightforward to show that
\begin{eqnarray}
[\,\hat{L}_{0i},\,\hat{P}_j\,]
& = & \pm\dfrac{i}{2}\left\{
(1-\hat{\mathbf{P}}^2)\delta_{ij}+2\hat{P}_i\hat{P}_j
\right\}
\;,
\cr
[\,\hat{L}_{ij},\,\hat{P}_k\,]
& = & 
i(\delta_{ik}\hat{P}_j - \delta_{jk}\hat{P}_i) \;,
\vphantom{\bigg|}
\end{eqnarray}
thus we can identify
\begin{equation}
\hat{X}_i \;=\; \pm\,2\hat{L}_{0i}\;.
\end{equation}

Next, consider a unit $D$-dimensional hyperboloid
in $(D+1)$-dimensional momentum space:
\begin{equation}
\hat{\eta}_0^2 - (\hat{\eta}_1^2+\hat{\eta}_2^2 + \cdots + \hat{\eta}_D^2) \;=\; 1\;.
\end{equation}
The generators which keep this hyperboloid invariant are
\begin{eqnarray}
\dfrac{1}{i}[\,\hat{M}_{\alpha\beta},\,\hat{\eta}_{\gamma}\,]
& = & g_{\alpha\gamma}\hat{\eta}_\beta - g_{\beta\gamma}\hat{\eta}_\alpha \;,
\cr
\dfrac{1}{i}[\,\hat{M}_{\alpha\beta},\,\hat{M}_{\gamma\delta}\,]
& = & g_{\alpha\gamma}\hat{M}_{\beta\delta} - g_{\alpha\delta}\hat{M}_{\beta\gamma}
    + g_{\beta\delta}\hat{M}_{\alpha\gamma} - g_{\beta\gamma}\hat{M}_{\alpha\delta}
\;,
\end{eqnarray}
where
\begin{equation}
g_{\alpha\beta} \;=\; \mathrm{diag}(-1,\underbrace{1,1,\cdots,1}_{\displaystyle n})\;.
\end{equation}
Writing $\hat{L}_{ij}=\hat{M}_{ij}$ and $\hat{K}_i=\hat{M}_{0i}$, we have
\begin{equation}
\begin{array}{rlrl}
{[\,\hat{L}_{ij},\,\hat{\eta}_{k}\,]}
& = \; i(\delta_{ik}\hat{\eta}_j - \delta_{jk}\hat{\eta}_i) \;,
& [\,\hat{K}_{i},\,\hat{\eta}_{j}\,]
& =\; -i\delta_{ij}\hat{\eta}_0 \;,
\vphantom{\Big|}\\
{[\,\hat{L}_{ij},\,\hat{\eta}_{0}\,]}
& = \;0 \;,
& [\,\hat{K}_{i},\,\hat{\eta}_{0}\,]
& =\; -i\hat{\eta}_i \;,
\vphantom{\Big|}\\
{[\,\hat{L}_{ij},\,\hat{L}_{k\ell}\,]}
& =\; i(\delta_{ik}\hat{L}_{j\ell} - \delta_{i\ell}\hat{L}_{jk} 
+ \delta_{j\ell}\hat{L}_{ik} - \delta_{jk}\hat{L}_{i\ell})
\;,
&&
\vphantom{\Big|}\\
{[\,\hat{L}_{ij},\,\hat{K}_{k}\,]}
& =\; i(\delta_{ik}\hat{K}_{j} - \delta_{jk}\hat{K}_{i})
\;,
&&
\vphantom{\Big|}\\
{[\,\hat{K}_{0i},\,\hat{K}_{0j}\,]}
& =\; -i\hat{L}_{ij} 
\;.
&&
\vphantom{\Big|}
\end{array}
\end{equation}
Stereographic projection from the North(South) pole (points $\eta_0=\pm 1$, $\eta_i=0$)
of the hyperboloid to the $\eta_0=0$ hyperplane yields 
\begin{equation}
\hat{P}_{i} \;=\; \dfrac{\hat{\eta}_i}{1\mp\hat{\eta}_0}\;,\qquad
\hat{\mathbf{P}}^2 
\;=\; \sum_{i=1}^{D}\hat{P}_{i}^2
\;=\; \dfrac{\hat{\eta}_0\pm 1}{\hat{\eta}_0\mp 1}
\;,
\label{PdefsHyper}
\end{equation}
the upper(lower) sign corresponding to the North(South)-pole projection.
See Fig.~\ref{Projection}(b).
Note that $\hat{\eta}_0^2 \ge 1$, which leads to the difference in sign for the
expression for $\hat{\mathbf{P}}^2$.
Again, it is straightforward to show that
\begin{eqnarray}
[\,\hat{K}_{i},\,\hat{P}_j\,]
& = & \pm\dfrac{i}{2}\left\{
(1+\hat{\mathbf{P}}^2)\delta_{ij}-2\hat{P}_i\hat{P}_j
\right\}
\;,
\cr
[\,\hat{L}_{ij},\,\hat{P}_k\,]
& = & 
i(\delta_{ik}\hat{P}_j - \delta_{jk}\hat{P}_i) \;,
\vphantom{\bigg|}
\end{eqnarray}
thus we can identify
\begin{equation}
\hat{X}_i \;=\; \pm\,2\hat{K}_{i}\;.
\end{equation}
Thus, our algebra can be understood in terms of a projection down from a higher dimension.

%%%%%%%%%%%%%%%%%%%%%%%%%%%%%%%%%%%%%%%%%%%%%%%%%%%%%%%%%%%%%%%%%%%%%%%%%%%%%%%%%%%%%%%%%%%%%%%%%
%%%%%%%%%%%%%%%%%%%%%%%%%%%%%%%%%%%%%%%%%%%%%%%%%%%%%%%%%%%%%%%%%%%%%%%%%%%%%%%%%%%%%%%%%%%%%%%%%
%\newpage
\bibliographystyle{ws-procs975x65}
\bibliography{MLUR}

\end{document}